\documentclass[conference,10pt]{IEEEtran}
\usepackage{cite}
\usepackage{amsmath,amssymb}
\usepackage{graphicx}
\usepackage{float}
\usepackage{booktabs,array}
\usepackage{url}
\newcommand{\E}{\mathbb{E}}
\newcommand{\R}{\mathbb{R}}
\newcommand{\C}{\mathbb{C}}
\newcommand{\cA}{\mathcal{A}}
\newcommand{\cC}{\mathcal{C}}
\newcommand{\cG}{\mathcal{G}}

\begin{document}
\title{GTAC: A Graph-Transformer Actor-Critic for\newline Carrier-Aware MU-MIMO User Grouping}
\author{\IEEEauthorblockN{Yan Huang\IEEEauthorrefmark{1},
Charles Santhosam Lourdu Raja\IEEEauthorrefmark{2},
Rajesh Gadiyar\IEEEauthorrefmark{1}}
\IEEEauthorblockA{\IEEEauthorrefmark{1}NVIDIA Corporation, Santa Clara, USA\\
\IEEEauthorrefmark{2}NVIDIA Corporation, Bangalore, India}}

\maketitle

\begin{abstract}
Achieving higher spectral efficiency (SE) in 5G and future 6G networks requires efficient massive multi-user multiple-input multiple-output (MU-MIMO) grouping under coupled physical-layer constraints and strict scheduling deadlines. 
Motivated by industry efforts in AI-native RAN, we propose Graph-Transformer Actor-Critic (GTAC), a deep reinforcement learning (DRL) architecture for carrier-aware MU-MIMO user grouping. The objective is to maximize the long-term proportional-fair (PF) utility of the cell subject to constraints from the practical stack protocol.
GTAC comprises a shared actor--critic encoder with spatial-edge-biased attention mechanism, a pointer decoder actor that generates grouping decisions, and a multi-stage transformer critic that estimates the carrier-wide state-value during training.
Training begins by cloning a teacher policy that cannot meet real-time deployment deadlines, followed by critic warmup and two-stage reinforcement learning using proximal policy optimization (PPO), first on a narrow, power-matched carrier and then on the full carrier with a Kullback--Leibler (KL) policy anchor.
After two-stage PPO, the best GTAC policy outperforms the non-real-time teacher by 11.6\% in carrier SE, while exceeding its PF utility and preserving the cell-edge throughput.
A CUDA implementation of GTAC achieves a 69.6~$\mu$s inference latency on an RTX PRO 4500 GPU, below the targeted 100~$\mu$s inference budget in practical systems.
\end{abstract}

\vspace{-6pt}
\section{Introduction}
Higher data traffic densities and user throughput requirements demand greater spectral efficiency (SE) in 5G/5G-Advanced and future 6G wireless systems~\cite{itu2160}. A key enabler is massive multiple-input multiple-output (MIMO), which provides substantial spatial multiplexing gains by allowing multiple user equipments (UEs) to share the same time-frequency resources. Realizing these gains requires forming spatially compatible and efficient multi-user (MU) groups across the frequency carrier.

Artificial intelligence and machine learning (AI/ML) offers a promising approach to this MU-MIMO user grouping optimization. 
ML policies can capture spatial compatibility among UEs and adapt grouping decisions to varying channel conditions, helping realize the SE potential of massive MU-MIMO. This use of AI/ML for radio optimization aligns with ongoing industry efforts on AI-native RAN, including the AI-RAN Alliance, SoftBank AITRAS, and NVIDIA Aerial~\cite{kundu2026airan}.

MU-MIMO user grouping is a combinatorial optimization problem with tightly coupled physical-layer (PHY) interactions. 
A grouping decision must balance beamforming gains, intra-group interference, and power sharing across UEs and spatial layers, while accounting for imperfect channel state information (CSI). 
Practical PHY transmission constraints further couple these decisions across the carrier. 
A key constraint is that a single modulation and coding scheme (MCS) must be applied across the physical resource block groups (PRBGs) and spatial layers carrying a UE's transport block~\cite{ts38214}. 
A weak PRBG can therefore reduce a UE's grant-wide goodput. Consequently, groups that appear favorable on individual PRBGs may be suboptimal when evaluated jointly across the carrier.

Beyond its complexity, MU-MIMO user grouping must satisfy strict real-time constraints. At 30-kHz subcarrier spacing, grouping shares a slot of duration $\Delta t=0.5$~ms with other scheduling and PHY functions, leaving only a fraction of the slot for grouping decisions. For the targeted scheduling pipeline, we impose a practical sub-100-$\mu$s latency requirement, $T_{\mathrm{grp}}<100~\mu\mathrm{s}<\Delta t$.
Existing solutions addressing these challenges can be broadly classified into three categories:

\noindent {\bf Optimization methods\/}
address MU-MIMO user grouping through optimization formulations and iterative algorithms that recompute beamforming weights and power allocation at each optimization step~\cite{bandi2020joint,choi2020joint}.
These methods yield high-quality solutions, but their cost grows with problem size and iteration count, making them unsuitable for real-time scheduling.

\noindent {\bf Heuristics designs\/}
reduce grouping complexity through spatial compatibility filtering and restricted search.
NVIDIA Aerial cuMAC provides a GPU-based semi-orthogonal MU-MIMO user grouping implemented for real-time scheduling~\cite{nvidia_aerial_cumac_source}.
Its design relies on spatial compatibility alone and cannot quantify goodput under the carrier-wide common-MCS coupling.
mCore+ combines channel-quality and chordal-distance filtering with multiphase parallel search for MU-MIMO scheduling~\cite{mcore2023}. However, its rate evaluation on each RB is confined to the retained candidate set and cannot reconsider discarded UEs that might form more favorable groups.
Overall, heuristic methods can discard favorable UE combinations and restrict solution space exploration due to fixed filtering rules.

\noindent {\bf Machine learning-based methods\/}
shift computational effort from online solution search to offline training, enabling scheduling decisions through ML model inference.
For MU-MIMO user grouping, Sun et al.~\cite{sun2026pnac} use a recurrent pointer actor with a Shannon sum-SE objective, whereas Zhu et al.~\cite{zhu2025transformer} employ CNN-transformer encoding for a narrowband uplink model.
Both approaches do not account for cross-PRBG common-MCS goodput coupling in their designs. 
Extending them to carrier-wide grouping therefore requires incorporating this coupling into policy training and value estimation.

\begin{samepage}
To address these challenges, we develop Graph-Transformer Actor-Critic (GTAC) for carrier-aware MU-MIMO user grouping. Our main contributions are as follows:
\begin{itemize}
\item The GTAC deep reinforcement learning (DRL) architecture for massive MU-MIMO user grouping, integrating a shared encoder, a pointer actor, and a multi-stage critic for carrier-wide state-value prediction.
\item A shared graph-transformer encoder with spatial edge biases that incorporate inter-UE MU-MIMO compatibility directly into the attention mechanism and the resulting aggregation of projected node values.
\item A multi-stage transformer critic that models MCS coupling across PRBGs and spatial layers per UE and explicitly incorporates available radio resources into carrier-wide state-value prediction.
\item A staged training procedure combining policy cloning from an offline teacher, critic warmup, and a two-stage PPO curriculum, first on a narrow, power-matched carrier and then on the full carrier with a Kullback--Leibler (KL) policy anchor, to learn grouping policies that exceed the teacher in SE without losing cell-edge throughput.
\item Performance evaluation showing that the best cloned policy retains 99.1\% of the teacher's SE, while the best two-stage PPO policy achieves 11.6\% SE gain over the teacher, with 104.1\% of the teacher's PF utility and its cell-edge throughput preserved.
A CUDA implementation of GTAC achieves median and p99 inference latencies of 67.6 and 69.6~$\mu$s on an RTX PRO 4500 GPU, meeting the targeted 100~$\mu$s latency budget in practical systems.
\end{itemize}
\end{samepage}


\section{Problem Modeling}
\label{sec:model}
\setlength{\textfloatsep}{6pt plus 2pt minus 2pt}
\setlength{\dbltextfloatsep}{6pt plus 2pt minus 2pt}
Consider the downlink of a cell where a BS with $N_t$ transmit antennas serves $U$ UEs, each equipped with $N_r$ receive antennas.
UEs are indexed by $k\in\mathcal K=\{1,\ldots,U\}$, with the connected UE set $\mathcal K$ assumed fixed over the scheduling horizon.
The system bandwidth consists of $G$ PRBGs, which are indexed by $g\in\{1,\ldots,G\}$.
Denote $B_g$ as the frequency bandwidth of PRBG $g$, and $B_{\mathrm{tot}}=\sum_{g=1}^{G}B_g$ as the total bandwidth of the carrier.
Time slots are indexed by $t$, and $\Delta t$ denotes the slot duration.
In each slot $t$, a time-domain scheduler (TDS) selects a candidate UE set $\cC_t\subseteq\mathcal K$ containing at most $K$ candidate UEs for scheduling. For example, a TDS may rank UEs based on proportional-fairness (PF) and shortlist the highest-priority candidates across the cell.

The reported rank $r_{tk}\in\mathcal L$ specifies the spatial-layer count of UE $k$ in slot $t$, where $\mathcal L\subseteq\{1,\ldots,\min(N_t,N_r)\}$. During MU-MIMO user grouping, each schduled UE's layer count is assumed to remain fixed at $r_{tk}$ across all its assigned PRBGs, subject to the NR layer and antenna-port mapping procedures~\cite[Secs.~7.3.1.3--7.3.1.5]{ts38211}.

For $k\in\cC_t$, the binary selection variable $x_{tgk}$ equals one if UE $k$ is scheduled on PRBG $g$ in slot $t$, and zero otherwise. Define the MU group on PRBG $g$ as $\cA_{tg}=\{k\in\cC_t:x_{tgk}=1\}$ and the carrier-wide schedule as $\cA_t=(\cA_{t1},\ldots,\cA_{tG})$. The selection variables satisfy $x_{tgk}\in\{0,1\}$ and $\sum_{k\in\cC_t}r_{tk}x_{tgk}\leq L_{\max}$ for every PRBG $g$ and candidate $k\in\cC_t$, where $L_{\max}$ is the maximum number of spatial layers supported per PRBG. Let $\mathcal F_t$ denote the set of carrier-wide schedules satisfying these constraints. 
A UE may be selected on multiple PRBGs and appears in different MU-MIMO groups~\cite{ts38214}.

Let $\mathbf H_{tgk},\widehat{\mathbf H}_{tgk}\in\C^{N_r\times N_t}$ denote the true downlink channels and their estimates from uplink sounding reference signals (SRS) under reciprocity. From the singular value decomposition $\widehat{\mathbf H}_{tgk}=\mathbf U_{tgk}\mathbf S_{tgk}\mathbf V_{tgk}^{H}$, stack the leading $r_{tk}$ rows of $\mathbf S_{tgk}\mathbf V_{tgk}^{H}$ for scheduled UEs into $\mathbf Z_{tg}\in\C^{L_{tg}\times N_t}$, where $L_{tg}=\sum_{k\in\cC_t}r_{tk}x_{tgk}$.
The eigen-regularized-zero-forcing (eigen-RZF) precoder is $\widetilde{\mathbf W}_{tg}=\mathbf Z_{tg}^{H}(\mathbf Z_{tg}\mathbf Z_{tg}^{H}+\alpha_{tg}\mathbf I_{L_{tg}})^{-1}$, with $\alpha_{tg}=\kappa_{\mathrm{RZF}}\operatorname{tr}(\mathbf Z_{tg}\mathbf Z_{tg}^{H})/L_{tg}$ and $\kappa_{\mathrm{RZF}}>0$. Its columns are unit-normalized, and each PRBG's transmit power is shared equally across its $L_{tg}$ layers.

Let $\gamma_{tgk\ell}\geq0$ denote the post-equalization signal-to-interference-plus-noise ratio (SINR) of spatial layer $\ell\in\{1,\ldots,r_{tk}\}$ for scheduled UE $k$ on PRBG $g$ in slot $t$. 

\noindent {\bf Carrier-Coupled PF Utility Objective\/} \ \ \
Denote $\cG_{tk}=\{g:x_{tgk}=1\}$ as the set of allocated PRBGs of a scheduled UE $k\in\cC_t$, and $\mathcal M$ as the set of feasible MCS levels. 
For each MCS $m\in\mathcal M$, UE $k$'s grant-wide effective SINR $\gamma^{\mathrm{eff}}_{tk}(m)$ can be modeled using exponential effective SINR mapping (EESM) as
\begin{equation}
 \gamma^{\mathrm{eff}}_{tk}(m)=-\beta_m\ln\!\left[\frac{\sum_{g\in\cG_{tk}}\sum_{\ell=1}^{r_{tk}}e^{-\gamma_{tgk\ell}/\beta_m}}{r_{tk}|\cG_{tk}|}\right],
 \label{eq:eesm}
\end{equation}
where $\beta_m>0$ is a MCS-specific EESM parameter.

A scheduled UE $k$ uses a single MCS $m_{tk}\in\mathcal M$ across its allocated PRBGs in $\cG_{tk}$. The NR TBS procedure~\cite{ts38214} maps allocated resources, rank, and MCS to transport-block size $D_{tk}$ in bits. Let $p_{tk}^{\mathrm{TB}}\in[0,1]$ be its decoding-error probability for one transmission attempt, which is a function of the effective SINR $\gamma^{\mathrm{eff}}_{tk}(m)$. Expected UE goodput $\overline R_{tk}$ and the weighted carrier-goodput objective $\eta_t$ are
\begin{equation}
\begin{aligned}
 \overline R_{tk}=\frac{D_{tk}}{\Delta t}(1-p_{tk}^{\mathrm{TB}}),\qquad
 \eta_t=\frac{1}{B_{\mathrm{tot}}}\sum_{k\in\cC_t}\varpi_{tk}\overline R_{tk},
\end{aligned}
 \label{eq:reward}
\end{equation}
where $\varpi_{tk}=R_{\mathrm{ref}}/\max\{\bar T_{tk},R_{\min}\}$ is UE $k$'s PF utility weight for slot $t$, $\bar T_{tk}$ is UE $k$'s average throughput before slot $t$, and  $R_{\mathrm{ref}},R_{\min}>0$.
In each slot $t$, carrier-wide MU-MIMO user grouping is formulated as
\begin{equation}
 \underset{\cA_t\in\mathcal F_t}{\operatorname{maximize}}\quad
 \eta_t(\cA_t),
 \label{eq:grouping-problem}
\end{equation}
where $\eta_t(\cA_t)$ is evaluated for schedule $\cA_t$.

\section{GTAC Actor-Critic DRL Architecture}
\label{sec:architecture}

\begin{figure}[t]
 \centering
 \includegraphics[width=0.90\columnwidth]{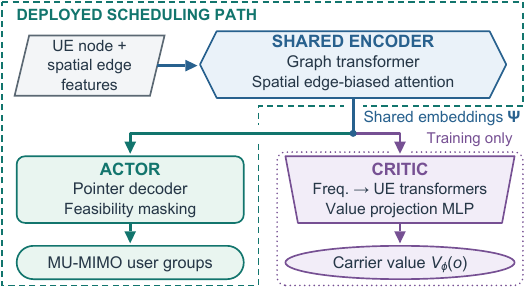}
 \caption{GTAC architecture with a shared graph-transformer encoder.}
 \label{fig:architecture-overview}
\end{figure}

The MU-MIMO grouping problem in~\eqref{eq:grouping-problem} has a combinatorial action space. 
The carrier-wide common-MCS coupling requires candidate schedules to be evaluated jointly across PRBGs. 
Repeatedly searching this space with detailed PHY evaluations is costly under strict scheduling deadlines. 
Further, each scheduling decision affects the data delivered to each UE in each slot and thus the long-term average throughput, which determines the PF utility in future slots. 
These computational and temporal dependencies motivate a DRL approach that learns grouping decisions from carrier-wide rewards while accounting for their impact on future scheduling performance. 

\noindent {\bf DRL Objective and Reward Design\/} \ \ 
GTAC learns a policy $\pi(\cA_t\mid o_t)$ that makes carrier-wide grouping decisions from observation $o_t$. 
For each step $t$, a reward $r_t$ is recieved (defined below). 
The DRL objective is
\begin{equation}
 J(\pi)=\E_{\pi}\!\left[\sum_{t=0}^{\infty}\gamma_{\mathrm{RL}}^{t}r_t\right],
 \label{eq:objective}
\end{equation}
where $0<\gamma_{\mathrm{RL}}<1$ is the discount factor, $\E_\pi$ averages over decisions and environment evolution. 
Maximizing the discounted return promotes higher future PF utility achieved by the DRL grouping.

The reward signal $r_t$ is designed to align with the PF grouping objective~\eqref{eq:reward}.
Since the decoding-error probabilities $p_{tk}^{\mathrm{TB}}$ are unavailable in practice, we use HARQ acknowledgments (ACKs) for the scheduled transimissions in the training process. 
Let $z_{tk}\in\{0,1\}$ indicate ACK of UE $k$'s transmission in slot $t$. The ACK-based PF-weighted carrier goodput is
\begin{equation}
 \eta_t^{\mathrm{ACK}}=
 \frac{\sum_{k\in\cC_t}\varpi_{tk}D_{tk}z_{tk}}{B_{\mathrm{tot}}\Delta t}.
 \label{eq:ack-reward}
\end{equation}
The acknowledged rate updates the PF average throughput of every connected UE $k\in\mathcal K$ as $\bar T_{t+1,k}=(1-\beta)\bar T_{tk}+\beta\frac{z_{tk}D_{tk}}{\Delta t}$, with $\beta\in(0,1]$.
Reward $r_t$ is defined as:
\begin{equation}
\begin{aligned}
 r_t&=c_r\big[\Phi(\bar{\mathbf T}_{t+1})-\Phi(\bar{\mathbf T}_t)\big],\\
 \Phi(\bar{\mathbf T})&=\frac{1}{|\mathcal K|}\sum_{k\in\mathcal K}
 \ln\frac{\max\{\bar T_k,\varepsilon\}}{T_{\mathrm{ref}}},
\end{aligned}
 \label{eq:pf_reward}
\end{equation}
where $\bar{\mathbf T}_t=(\bar T_{tk})_{k\in\mathcal K}$ collects the UE average throughput, $\varepsilon>0$ keeps the logarithm finite, and $c_r>0$ scales the reward. The fixed reference rate $T_{\mathrm{ref}}>0$ makes the logarithm dimensionless and cancels in the utility difference.

When the relative changes in average throughput are small, a first-order expansion gives $r_t\approx c_r\beta\big(\frac{B_{\mathrm{tot}}\eta_t^{\mathrm{ACK}}}{|\mathcal K|R_{\mathrm{ref}}}-1\big)$. Thus, a higher expected slot reward approximately corresponds to a higher PF-weighted grouping objective~\eqref{eq:reward}.


\noindent {\bf Shared Graph-Transformer Encoder\/} \ \ 
GTAC shares a graph-transformer encoder between actor and critic (Fig.~\ref{fig:architecture-overview}) because user grouping and carrier-wide value prediction both rely on channel quality and inter-UE spatial compatibility.  
The actor and critic use separate decoders for grouping decisions and carrier-wide state-value estimation, respectively.
Let $\theta_E$, $\theta_A$, and $\theta_C$ denote the parameters of the shared encoder, actor decoder, and critic decoder, respectively. $\theta=(\theta_E,\theta_A)$ and $\phi=(\theta_E,\theta_C)$ denote the complete actor and critic parameters.

For each PRBG $g$, the candidate UEs in $\cC_t$ form a \emph{fully-connected graph} whose nodes represent UEs and whose edges encode inter-UE spatial compatibility. The encoder combines UE features and their spatial relationships into shared embeddings that support the actor's grouping decisions and the critic's carrier-wide value estimation.

In the per-PRBG graph, each node $k\in\cC_t$ has feature vector $\mathbf f_{gk}=[\mathbf b_{gk},\mathbf f^{\mathrm{ch}}_{gk},\mathbf f^{\mathrm{st}}_{gk},\mathbf f^{\mathrm{ctx}}_g]$, with the respective input dimensions $F_b,F_{c},F_{s},F_g$.
Beam features $\mathbf b_{gk}$ encode normalized powers over a discrete Fourier transform (DFT) beam dictionary available at the BS for CSI processing~\cite[Sec.~5.2.2.2.1]{ts38214}.
CSI-derived channel, spatial, and predicted rate features $\mathbf f^{\mathrm{ch}}_{gk}$ summarize channel-mode strengths, rank validity, angular and polarization structure, and channel quality indicator (CQI)-based rate proxies with PF weighting.
CSI and scheduling-state features $\mathbf f^{\mathrm{st}}_{gk}$ capture reported CQI and rank, report and scheduling ages, mean throughput, and power and precoder regularization levels.
PRBG-context features $\mathbf f^{\mathrm{ctx}}_g$, shared across candidate UEs, describe relative frequency position, power/noise conditions, and candidate UE occupancy.

\begin{figure*}[t]
 \centering
 \includegraphics[width=0.82\textwidth]{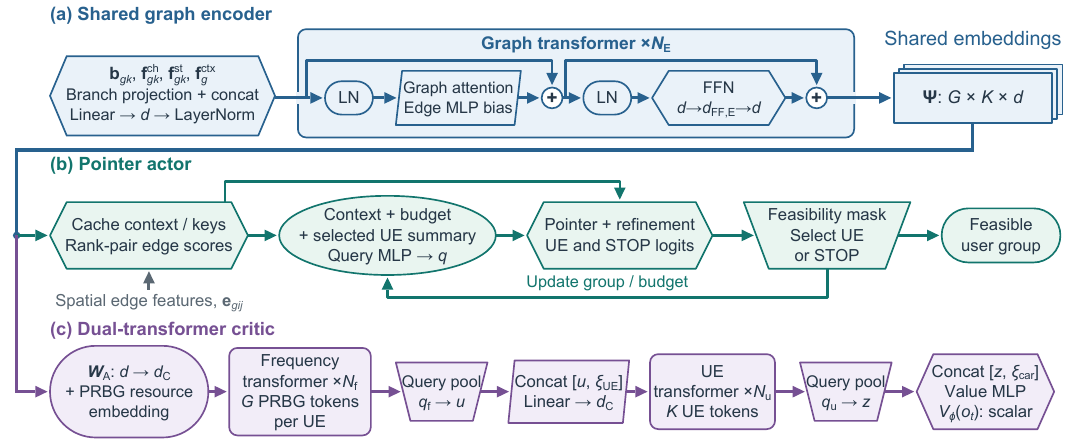}
 \caption{GTAC architecture. Transformer depths: $N_E,N_f,N_u$. Critic width: $d_C$. Edge features $\mathbf e_{gij}$ describe spatial compatibility of UEs $i, j$ on PRBG $g$.}
 \label{fig:architecture}
\end{figure*}

Edge $(i,j)$ has feature tensor $\mathbf e_{gij}\in\R^{|\mathcal L|\times|\mathcal L|\times d_e}$, with $d_e$ features per supported rank pair. These features summarize inter-UE channel correlations, relative powers and ranks, and DFT-beam overlap and polarization compatibility. Detailed node and edge feature design specifications are omitted from this paper due to space limitations.

The four encoder branches in Fig.~\ref{fig:architecture}(a) each apply an affine projection, with output widths $d_b,d_r,d_c,d_g$, respectively. Their outputs are concatenated to width $d_\Sigma=d_b+d_r+d_c+d_g$, projected to the shared embedding width $d$.

For standard query/key/value projections, the attention mechanism, and value aggregation, see~\cite[Sec.~3.2]{vaswani2017attention}. Within one encoder block, let $\boldsymbol\psi_{gi}^{\mathrm{in}}\in\R^d$ denote the input embedding and $\mathbf z_{gi}=\operatorname{LN}_1(\boldsymbol\psi_{gi}^{\mathrm{in}})$ its layer-normalized version. Spatial relations bias scaled dot-product attention for each of the $n_h$ attention heads, with head width $d_h=d/n_h$:

\begin{equation}
 a_{gij}^{h}=\frac{(\mathbf Q_h\mathbf z_{gi})^T
 (\mathbf K_h\mathbf z_{gj})}{\sqrt{d_h}}
 +b_h\!\left(\operatorname{vec}(\mathbf e_{gij})\right).
 \label{eq:attention}
\end{equation}
An edge-bias multilayer perceptron (MLP) maps the flattened edge features of dimension $d_{\mathrm{flat}}=|\mathcal L|^2d_e$ through a hidden layer of width $d_{\mathrm{bias}}$ to $n_h$ scalar outputs, one attention bias $b_h$ per head $h$. Softmax over $\cC_t$ weights projected values:
\begin{equation}
 \omega_{gij}^{h}=\frac{\exp(a_{gij}^{h})}{\sum_{k\in\cC_t}\exp(a_{gik}^{h})},
 \qquad
 \mathbf y_{gi}^{h}\in\R^{d_h}=\sum_{j\in\cC_t}\omega_{gij}^{h}\mathbf V_h\mathbf z_{gj}.
\label{eq:edge_aggregation}
\end{equation}
Each head output $\mathbf y_{gi}^{h}$ aggregates candidate-UE information for UE $i$ on PRBG $g$. Concatenating the heads gives $\mathbf y_{gi}=\operatorname{Concat}(\mathbf y_{gi}^{1},\ldots,\mathbf y_{gi}^{n_h})\in\R^d$. The output projection and two residual updates in Fig.~\ref{fig:architecture}(a) are \mbox{$\boldsymbol\psi_{gi}^{\mathrm{attn}}=\boldsymbol\psi_{gi}^{\mathrm{in}}+\mathbf W_O\mathbf y_{gi}+\mathbf b_O$} and \mbox{$\boldsymbol\psi_{gi}^{\mathrm{out}}=\boldsymbol\psi_{gi}^{\mathrm{attn}}+\operatorname{FFN}\!\left(\operatorname{LN}_2(\boldsymbol\psi_{gi}^{\mathrm{attn}})\right)$}, where $\mathbf W_O\in\R^{d\times d}$ and $\mathbf b_O\in\R^d$ define the learned output projection, $\operatorname{LN}_1$ and $\operatorname{LN}_2$ are separate layer normalizations, and FFN denotes the feedforward network. Each FFN has hidden width $d_{\mathrm{FF},E}$.

With fixed queries and keys, larger attention edge biases emphasize compatible UEs or strong interferers, capturing co-scheduling interactions without costly pairwise precoder solves.
This graph-transformer encoder adapts structural attention biases~\cite{ying2021graphormer} to model spatial relationships among UEs.

The encoder stacks $N_E$ such graph-transformer blocks. The final block outputs form the shared tensor $\boldsymbol\Psi\in\R^{G\times K\times d}$, which stacks embeddings $\boldsymbol\psi_{gk}\in\R^d$ for UE $k$ on PRBG $g$ and supplies both actor and critic.

\noindent {\bf Actor: Feasibility-Masked Pointer Decoder\/} \ \ 
The actor constructs a MU-MIMO group on each PRBG by selecting UEs sequentially. 
A pointer decoder assigns probabilities directly to candidate UEs and a learned STOP action.

At decoding step $j$, let $\cA_{tg}^{(<j)}$ denote the selected UE set. Learned attention pooling summarizes the candidate embeddings into a PRBG context vector $\mathbf c_g$. A query MLP combines this context with the rank-weighted mean of selected-UE embeddings, the remaining budget, decoding progress, and selected-group statistics to produce $\mathbf q_{gj}\in\R^d$.
Candidate UE $k$ receives the score:
\begin{equation}
 s_{gjk}
 =\frac{(\mathbf W_p\boldsymbol\psi_{gk})^T\mathbf q_{gj}}{\sqrt d}
 +f_{\mathrm{ref}}(\mathbf v_{gjk}),
 \label{eq:pointer}
\end{equation}
where $\mathbf W_p$ is a learned key projection and $f_{\mathrm{ref}}$ is a scalar-valued refinement MLP.
The first term matches each encoded candidate to the current grouping context. 
The refinement $f_{\mathrm{ref}}$ explicitly evaluates its relationship with the selected group. Its input $\mathbf v_{gjk}$ combines candidate attributes, group statistics, the budget after admission, and aggregated compatibility with selected UEs. A separate edge MLP maps spatial features at the reported rank pairs to compatibility scores, which are cached and reused throughout decoding. Thus, the encoder represents relationships across all candidates, while refinement makes compatibility with the particular group under construction explicit. These learned corrections avoid repeated precoder and SINR computations.

Feasibility is enforced directly through the admissible candidate set $\mathcal V_{tgj}=\{k\in\cC_t\setminus\cA_{tg}^{(<j)}:r_{tk}\leq L_{\max}-\sum_{u\in\cA_{tg}^{(<j)}}r_{tu}\}$.
Candidate and STOP scores are clipped using $\tau\tanh(s/\tau)$, with $\tau>0$, before invalid candidate logits are set to $-\infty$. Softmax assigns zero to infeasible selections.

A separate MLP computes the STOP score from $[\mathbf q_{gj},\mathbf c_g]$, allowing the policy to terminate before exhausting the layer budget. 
Each admission updates the selected-group summary and budget for the next decision, providing autoregressive conditioning without recurrent neural-network layers. PPO rollouts sample actions, whereas deployment selects the highest-probability action, with at most $T_{\max}$ decoding steps per PRBG. PRBGs can be decoded in parallel, while carrier-wide training rewards provide feedback on their coupled common-MCS outcomes.

\noindent {\bf Critic: Dual-Transformer Value Decoder\/} \ \ 
The critic estimates the discounted return of the carrier-wide reward from the shared encoder's output embeddings. As shown in Fig.~\ref{fig:architecture}(c), two transformer stages with learned-query pooling aggregate information across PRBGs and UEs.

For each candidate UE $k\in\cC_t$, the frequency stage attends across all $G$ PRBGs. Let $R_g$ be the available data resource elements (REs) on PRBG $g$, with fixed resource, PRBG-count, and candidate-count reference scales $R_0,G_0,K_0$. A frequency transformer $T_f$ and learned-query pooling $P_f$ process resource-augmented embeddings:
\begin{equation}
 \widetilde{\mathbf u}_k=\mathbf W_c\!\bigl[
 P_f\!\bigl(T_f(\{\mathbf W_A\boldsymbol\psi_{gk}
 +\mathbf W_R\tfrac{R_gG_0}{R_0}\}_{g=1}^{G})\bigr),
 \tfrac{G}{G_0},\tfrac{\sum_g R_g}{R_0}\bigr].
 \label{eq:freq}
\end{equation}
Here $\mathbf W_A\in\R^{d_C\times d}$ adapts the encoder width $d$ to the critic width $d_C$.
The UE stage aggregates candidate UEs through transformer $T_u$ and pooling $P_u$:
\begin{equation}
 V_\phi(o_t)=f_V\!\left[ P_u(T_u(\{\widetilde{\mathbf u}_k\}_{k\in\cC_t})),
 \frac{|\cC_t|}{K_0},\frac{\sum_gR_g}{R_0}\right].
 \label{eq:value}
\end{equation}
The stages use depths $N_f,N_u$, distinct weights, width $d_C$, $n_h$ heads, and FFN width $d_{\mathrm{FF},C}$. Their pre-LN residual structure follows the encoder without edge biases.
The appended count/resource vectors $\boldsymbol\xi_{\mathrm{UE}},\boldsymbol\xi_{\mathrm{car}}$ have width $d_\xi$. Pooling uses separate learned queries, bias-free key/value maps, and output projections, without LN, FFNs, or residual connections. A $(d_C+d_\xi)\to d_C\to\mathrm{scalar}$ MLP produces the signed value.

The critic first aggregates each UE's embeddings across PRBGs, then models competition among UEs to estimate carrier-wide value under the common-MCS coupling across each UE's allocated PRBGs. This value provides the baseline for a PPO advantage shared across PRBGs, guiding actor refinement using PF-utility feedback from the slot schedule.

\noindent {\bf Staged Training Framework\/} \ \ 
GTAC training progresses from imitation of an offline teacher to carrier-wide value estimation and two-stage policy refinement through PPO~\cite{schulman2017ppo}:

\begin{itemize}
\item \emph{Policy cloning:} Teacher-forced log-likelihood training initializes the encoder and actor from teacher solutions. The cloning checkpoint with the highest PF utility in the evaluation of Section~\ref{sec:evaluation} starts the next stages.
\item \emph{Critic warmup:} The cloned policy first runs for $N_b$ slots to stabilize UE throughput averages. Then $\theta_C$ minimizes mean squared error against carrier returns with $\theta_E,\theta_A$ frozen. The same preparation under the frozen actor precedes each stage-specific critic re-warmup.
\item \emph{Two-stage PPO:} Stage-1 PPO runs on a narrow carrier of $G'<G$ PRBGs whose BS and UE transmit powers are scaled to preserve per-PRBG transmit power relative to the full carrier. Each PRBG decision then weighs more in the carrier-level advantage shared across PRBGs. Stage-1 checkpoints are evaluated on the full carrier, and the one with the highest PF utility starts stage 2. Stage-2 PPO continues on the full carrier. The full-carrier critic is re-warmed under the selected stage-1 actor. 
A KL anchor ties the policy to that actor. The learning rate ramps linearly over the first $N_{\mathrm{ramp}}$ updates.
\end{itemize}

The teacher is a PF/eigen-RZF greedy algorithm. On each PRBG, it initializes a group from PF-ranked candidates and iteratively adds feasible UE with the largest PF-utility gain. Selection stops when no improving addition remains or layer budget is met. Each candidate addition requires recomputing the precoder and SINRs with the new UEs, making the teacher compute-intensive and unsuitable for real-time scheduling.

Critic warmup and PPO use the carrier-wide reward in~\eqref{eq:pf_reward}.
Generalized advantage estimation (GAE) with parameter $\lambda$ supplies carrier advantages and value targets~\cite[Sec.~5]{schulman2017ppo}. Value loss $L_V$ uses the larger squared error of clipped and unclipped value predictions.

Given the observation $o_t$, the actor generates the grouping sequence for each PRBG independently.
For ordered sequence $A_{tg}$ and PRBG observation $o_{tg}$, $\log\pi_\theta(A_{tg}\mid o_{tg})=\sum_{j\in\mathcal J_{tg}}\log\pi_\theta(a_{tgj}\mid a_{tg,<j},o_{tg})$, where $\mathcal J_{tg}$ includes UE selections and a learned STOP.
The policy loss $L_\pi$ is the negative mean PPO clipped surrogate~\cite{schulman2017ppo} over valid PRBG sequences. Each uses the ratio $\rho_{tg}=\pi_\theta(A_{tg}\mid o_{tg})/\pi_{\mathrm{old}}(A_{tg}\mid o_{tg})$ and the shared normalized carrier advantage $\widehat A_t$, with clipping radius $\epsilon$ applied separately to each PRBG ratio.

\begin{table}[!t]
\caption{Simulation and training configuration}
\label{tab:simulation}
\centering\fontsize{7.5}{8.5}\selectfont
\begin{tabular}{@{}p{.43\columnwidth}p{.52\columnwidth}@{}}
\toprule
Parameter & Configuration\\\midrule
Scenario / carrier & 3GPP TR 38.901 UMa / 3.5 GHz\\
Subcarrier spacing / slot & 30 kHz / 0.5 ms\\
Bandwidth / PRBG & 100 MHz / 17 PRBGs (16 PRBs/PRBG)\\
BS / UE antenna ports & $N_t=4\times8\times2=64$ / $N_r=2$\\
Transmit power / noise figure & 46 dBm / 7 dB\\
Active UE population / UE speed & 128 connected UEs per sector / 3 km/h\\
TDS candidates / layer limit & $K=32$ / $L_{\max}=16$\\
CSI report periodicity & 20 ms (40 slots)\\
CSI error model & SINR-scaled, $-15$ dB NMSE for SRS\\
Allocation / traffic & RAT-0 / full buffer\\
Precoder / link adaptation & Eigen-RZF / genie-aided\\\midrule
Steps: Clone / critic warmup / PPO & 2000 / 500 / 400, 400  for PPO stage 1 and 2\\
Learning rates: clone / PPO encoder and actor / PPO critic & $(3,1,3)\times10^{-4}$\\
Rollout slots / minibatch slots & 16 / 32\\
PPO epochs / $T_{\max}$ & 4 / 16\\
$(\gamma_{\mathrm{RL}},\lambda,\epsilon)$ & $(0.99,0.95,0.2)$\\
$(c_V,c_H)$ & $(0.5,0.01)$\\
PF reward $(\beta,\varepsilon,c_r)$ & $(10^{-3},1~\mathrm{kbit/s},10^3)$\\
Throughput stabilization $N_b$ & 3000 slots\\
Stage-1 carrier $G'$ / power scaling & 4 PRBGs / $-6.3$ dB (BS and UE)\\
Stage-2 $c_{\mathrm{KL}}$ / $N_{\mathrm{ramp}}$ & 0.05 / 20 steps\\
$(\tau,\kappa_{\mathrm{RZF}},\delta_{\mathrm{KL}})$ & $(10,0.01,0.03)$\\\bottomrule
\end{tabular}
\end{table}

In stage~2, the surrogate uses the anchored advantage
\begin{equation}
 \widetilde A_{tg}=\widehat A_t-c_{\mathrm{KL}}
 \ln\frac{\pi_{\mathrm{old}}(A_{tg}\mid o_{tg})}{\pi_{\mathrm{anc}}(A_{tg}\mid o_{tg})},
 \label{eq:kl_anchor}
\end{equation}
where $\pi_{\mathrm{anc}}$ is the frozen stage-1 policy and $c_{\mathrm{KL}}>0$. The penalty's expectation under $\pi_{\mathrm{old}}$ is $c_{\mathrm{KL}}D_{\mathrm{KL}}(\pi_{\mathrm{old}}\,\|\,\pi_{\mathrm{anc}})$, discouraging drift from that policy.

We minimize $L=L_\pi-c_H\mathcal H+c_VL_V$, with entropy $\mathcal H$ averaged over valid steps within each sequence and then over valid sequences, and value loss counted once per slot. Minibatches recompute outputs from stored observations with rollout statistics and targets fixed. One optimizer updates all parameter groups. A stop-gradient at $\boldsymbol\Psi$ confines $L_V$ gradients to $\theta_C$, while the policy and entropy terms train $\theta_E,\theta_A$. 

\begin{figure*}[t]
\centering
\includegraphics[width=\textwidth]{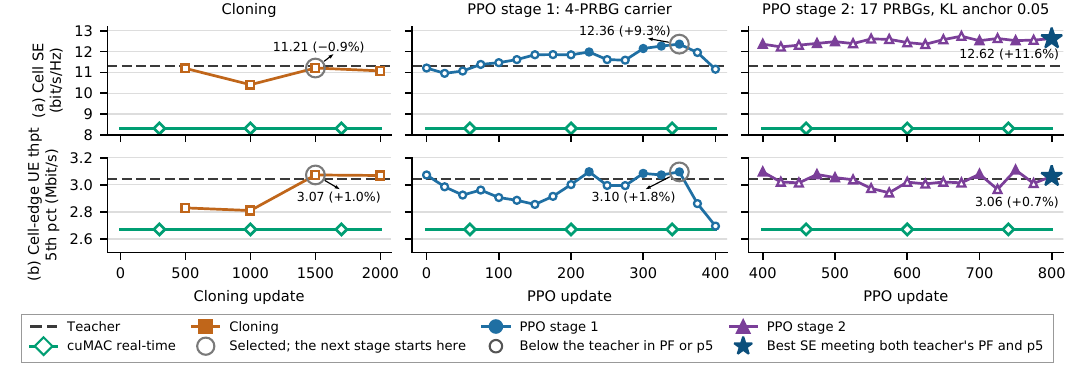}
\caption{Performance of GTAC training checkpoints. (a) Carrier SE. (b) Cell-edge 5th-percentile UE throughput. The star marks the highest-SE checkpoint whose PF utility and cell-edge throughput both reach the teacher's. Selected checkpoints are labeled with metric values and percentage gains over the teacher.}
\label{fig:checkpoint-progress}
\end{figure*}

\section{Performance Evaluation}
\label{sec:evaluation}

Table~\ref{tab:simulation} summarizes the simulation and training configurations.
Experiments run with 72 Arm Neoverse-V2 CPU cores in a single socket, one thread per core, and a maximum clock frequency of 3.384~GHz.
GPU implementations use an NVIDIA RTX PRO 4500 Blackwell Server Edition GPU with 82 streaming multiprocessors (SMs).

We evaluated GTAC across transformer depths $N_E,N_f,N_u$ and embedding widths $d$.  
Results below use the best-performing option with $(d,N_E,n_h)=(32,2,4)$, $N_f=N_u=1$, and critic width $d_C=64$.
The baselines are the offline PF/eigen-RZF teacher and Aerial cuMAC's semi-orthogonality-based user grouping~\cite{nvidia_aerial_cumac_source}, representing a widely used real-time heuristic design.
All schemes are evaluated on the full 100 MHz carrier (17 PRBGs). 
We report carrier SE, i.e.,~\eqref{eq:ack-reward} with unit weights, the PF utility, and the cell-edge throughput represented by the 5th-percentile of $T_k$ over $k\in\mathcal K$.

\noindent {\bf SE, PF Utility, and Cell-Edge\/} \ \ \
The teacher and cuMAC achieve SEs of 11.31 and 8.31~bit/s/Hz and cell-edge throughputs of 3.04 and 2.67~Mbit/s, respectively. 
cuMAC retains 85.8\% of the teacher's PF utility. The selected clone reaches 11.21~bit/s/Hz, 3.07~Mbit/s, and 99.6\% of the teacher's PF utility. Stage 1 (update 350) reaches 12.36~bit/s/Hz and 3.10~Mbit/s. Stage 2 (update 800; Fig.~\ref{fig:checkpoint-progress}) achieves 12.62~bit/s/Hz and 3.06~Mbit/s, with 104.1\% of the teacher's PF utility. Its SE exceeds the teacher, cuMAC, and clone by 11.6\%, 51.9\%, and 12.6\%, respectively. Update 675 achieves the highest SE (12.74~bit/s/Hz) but is excluded because its cell-edge throughput (3.01~Mbit/s) falls below the teacher's.

\noindent {\bf CPU and GPU Inference Latency\/} \ \ \
Table~\ref{tab:latency} reports GTAC encoder--actor inference and teacher latencies for one cell with $G=17$ and $K=32$. 
The CPU actor uses FP32 Inductor, autotuned GEMMs, and pinned, spinning workers ($17\times3$ threads plus a coordinator, 52 cores). 
GPU inference employs CUDA Graph with three fused CUDA kernels: node features/projections, rank-conditioned edges/biases, and graph-transformer/greedy decoding. 
CUDA implementation utilizes FP16 tensor core multiplications with FP32 accumulation.

GTAC achieves median/p99 encoder--actor inference latencies of 67.6/69.6~$\mu$s on the GPU, compared with 4.2/4.9~ms on the CPU. GPU inference provides median and p99 speedups of $62.1\times$ and $70.4\times$, respectively.

\begin{table}[!t]
\caption{Latency ($\mu$s): GTAC encoder--actor inference and teacher.}
\label{tab:latency}
\centering\footnotesize
\setlength{\tabcolsep}{4pt}
\begin{tabular}{@{}llrr@{}}
\toprule
Processor & Policy & Median & p99\\\midrule
Neoverse-V2 CPU & GTAC actor, 52 cores & 4\,200 & 4\,900\\
 & Teacher, 35 cores & 46\,100 & 56\,500\\\midrule
RTX PRO 4500 & GTAC actor & 67.6 & 69.6\\
Blackwell GPU & Teacher & 10\,200 & 12\,800\\\bottomrule
\end{tabular}
\end{table}

\section{Conclusions}
\label{sec:conclusions}
This paper presents GTAC, an actor--critic DRL architecture for carrier-aware massive MU-MIMO user grouping. It combines a shared graph-transformer encoder with spatial-biased attentions, a pointer actor, and a dual-transformer critic.
Cloned GTAC policy achieves 99.1\% of the SE of a computationally intensive, non-real-time teacher. Two-stage PPO raises SE to 11.6\% above the teacher, with higher PF utility and preserved cell-edge throughput.
A CUDA implementation of GTAC achieves 69.6~$\mu$s p99 inference latency, meeting the targeted 100~$\mu$s per-slot latency budget in practical systems.

\bibliographystyle{IEEEtran}
\bibliography{references}
\end{document}